\documentclass[journal]{IEEEtran}

\ifCLASSINFOpdf
\else
   \usepackage[dvips]{graphicx}
\fi
\usepackage{url}
\usepackage{graphicx}
\usepackage{enumerate}
\usepackage{breqn}
\usepackage{algorithm,cite}
\usepackage[noend]{algpseudocode}
\usepackage{multirow}
\usepackage{soul,color}
\usepackage{balance}
\usepackage{amsmath,amsthm,amssymb}
\usepackage{hyperref}

\begin{document}

\title{Cyclic Defense GAN Against Speech Adversarial Attacks}

\author{Mohammad Esmaeilpour,~\IEEEmembership{Member,~IEEE,}
 Patrick Cardinal,~\IEEEmembership{Member,~IEEE,}
 and~Alessandro Lameiras Koerich,~\IEEEmembership{Member,~IEEE}
\thanks{M. Esmaeilpour, P. Cardinal, and A. L. Koerich are with the \'{E}cole de Technologie Sup\'{e}rieure (\'{E}TS), Universit\'{e} du Qu\'{e}bec, Montr\'{e}al, QC, Canada,
contact: mohammad.esmaeilpour.1@ens.etsmtl.ca.

This work was funded by the NSERC under grants RGPIN 2016-04855 and 2016-06628. Supplementary material and source codes are available at \href{https:// github.com/EsmaeilpourMohammad/CD-GAN.git.}{this github-Repo.}}}

\markboth{IEEE Signal Processing Letters, August 2021}
{Shell \MakeLowercase{\textit{et al.}}: Bare Demo of IEEEtran.cls for IEEE Journals}
\maketitle

\begin{abstract}
This paper proposes a new defense approach for counteracting state-of-the-art white and black-box adversarial attack algorithms. Our approach fits into the implicit reactive defense algorithm category since it does not directly manipulate the potentially malicious input signals. Instead, it reconstructs a similar signal with a synthesized spectrogram using a cyclic generative adversarial network. This cyclic framework helps to yield a stable generative model. Finally, we feed the reconstructed signal into the speech-to-text model for transcription. The conducted experiments on targeted and non-targeted adversarial attacks developed for attacking DeepSpeech, Kaldi, and Lingvo models demonstrate the proposed defense's effectiveness in adverse scenarios.
\end{abstract}


\IEEEpeerreviewmaketitle

\section{Introduction}
\label{sec:intro}
There is a relatively increasing volume of publications on developing adversarial attacks against speech-to-text (transcription) systems in targeted and non-targeted scenarios \cite{carlini2018audio,qin2019imperceptible,chen2020metamorph,schonherr2020imperio,esmaeilpour2021towards}. These attack algorithms' effectiveness has been demonstrated for DeepSpeech \cite{MozillaImplementation}, Kaldi \cite{povey2011kaldi}, and Lingvo \cite{shen2019lingvo} transcription systems. In general, these adversarial attacks run an optimization algorithm for $\left \langle \vec{x}_{\mathrm{orig}},\hat{\mathbf{y}}_{i} \right \rangle$, where $\vec{x}_{\mathrm{orig}}$ stands for the original (legitimate) speech signal, and $\hat{\mathbf{y}}_{i}$ indicates the associated target phrase defined by the adversary~\cite{carlini2018audio}:  
\begin{dmath}
    {\min_{\delta} \left \| \delta \right \|_{F} + \sum_{i}c_{i}L_{i}(\vec{x}_{\mathrm{adv}},\hat{\mathbf{y}}_{i}) \quad \mathrm{s.t.} \quad l_{\text{dB}}(\vec{x}_{\mathrm{adv}}) < \epsilon}
    \label{eq:generalEqu}
\end{dmath}
\noindent where $l_{\text{dB}}(\cdot)$ denotes the loudness metric. Additionally, $\vec{x}_{\mathrm{adv}}=\vec{x}_{\mathrm{orig}}+\delta$ and $\delta$ denotes the adversarial perturbation achievable through this iterative optimization formulation. Moreover, $c_{i}$ is an hyperparameter for scaling the loss function $L_{i}(\cdot)$ by considering the length of the ground truth phrase $\mathbf{y}_{i}$ ($\mathbf{y}_{i}\neq \hat{\mathbf{y}}_{i}$). Furthermore, $l_{\text{dB}}(\cdot)$ computes the relative loudness (the distortion condition) of the signal in the logarithmic $\mathrm{dB}$-scale, and $\epsilon$ is the audible threshold defined by the adversary. There are several variants for Eq.~\ref{eq:generalEqu} where they often employ different loss functions, distortion conditions, and expectation over transformations (EOT). Carlini {\it et al.}~\cite{carlini2018audio} introduced the baseline variant of the aforementioned adversarial optimization formulation (C\&W attack), which incorporates the connectionist temporal classification (CTC) loss function $L_{i}(\cdot) \leftarrow \mathcal{L}_{i}(\cdot)$ \cite{graves2006connectionist}. 
This white-box attack is targeted, and it has been successfully characterized against the DeepSpeech transcription system. However, this algorithm is not robust against over-the-air playbacks, and it might simply bypass the optimized adversarial perturbation $\delta$ after replaying $\vec{x}_{\mathrm{adv}}$ over a noisy environment \cite{carlini2018audio,yakura2018robust}.

Yakura {\it et al.}~\cite{yakura2018robust} introduced a second variant of Eq.~\ref{eq:generalEqu}. They proposed an EOT operation to tackle the over-the-air playback issue. This operation implements the room impulse response (RIR) filter set and extends Eq.~\ref{eq:generalEqu} to \cite{yakura2018robust}:
\begin{equation}
    \min_{\delta} \mathbb{E}_{t \in \tau, \omega}\left [ \mathcal{L} ( \mathrm{mfcc}(\vec{x}_{\mathrm{adv}}),\hat{\mathbf{y}}_{i})+\alpha_{t}\left \| \delta \right \| \right ] 
    \label{eq:yakuraatt}
\end{equation}
\noindent where $\alpha_{t}$ is a scalar for adjusting the adversarial perturbation, $\tau$ denotes the EOT filter set including room impulse response, and $\omega$ is a white Gaussian noise filtration operation. Both $t$ and $\omega$ parameters contribute to capturing environmental distributions w.r.t. enclosed room settings. Additionally, $\mathrm{mfcc}(\cdot)$ refers to the standard Mel-frequency cepstral coefficient transform \cite{davis1980comparison}. This white-box attack algorithm yields:
\begin{equation}
    \vec{x}_{\mathrm{adv}} \leftarrow  \left [ \vec{x}_{\mathrm{orig}}+\Omega(\delta) \right ]\circledast t+\omega
\end{equation}
\noindent where $\circledast$ is the convolution operator, and $\Omega(\cdot)$ indicates the band-pass filtration operation for limiting the perturbation between 1 kHz and 4 kHz. Like the C\&W attack, the Yakura attack also uses the CTC loss function with a different distortion condition ($\left \| \delta \right \|< \epsilon$) and EOT operation. The results on attacking the DeepSpeech model corroborate the higher capacity of such an adversarial algorithm than the C\&W attack \cite{yakura2018robust}.

Sch{\"o}nherr {\it et al.}~\cite{schonherr2020imperio} introduced the Imperio attack, which is the third variant of Eq.~\ref{eq:generalEqu}. They presented a more straightforward simulation procedure for implementing the EOT operation in a noisy environment, which essentially fits in the targeted scenario within the white-box framework. The EOT operation incorporated in the Imperio attack is adapted for transcription models using conventional learning blocks such as a hidden Markov model in the Kaldi system: 
\begin{equation}
    \vec{x}_{\mathrm{adv}}  = \arg \max_{\vec{x}_{i}}\mathbb{E}_{t\sim \tau_{d}}\left [ P(\hat{\mathbf{y}}_{i}|\vec{x}_{i,t}) \right ]
    \label{eq:imperioatt}
\end{equation}
\noindent where $\tau_{d}$ is an RIR filter set with adequately large dimension $d$, and $P(\cdot)$ denotes the logits of a deep neural network (DNN) used for decoding $\hat{\mathbf{y}}_{i}$. This attack's distortion condition is $\left \| \delta \right \| < \epsilon_{p}$, where $\epsilon_{p}$ is a psychoacoustic thresholding, and the employed loss function is the cross-entropy ($\ell_{net} (\cdot)$) \cite{schonherr2020imperio}. 
\begin{figure*}[h]
  \centering
  \includegraphics[width=0.8\textwidth]{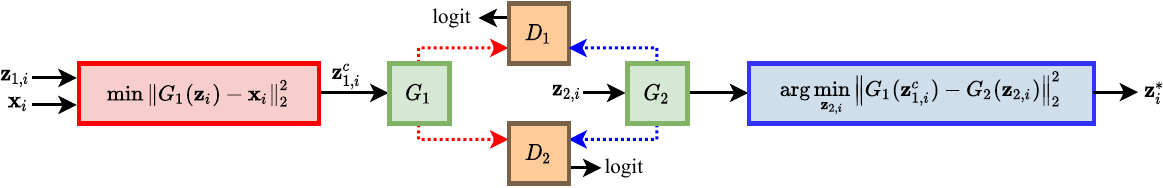}
  \caption{Overview of the proposed safe vector optimization procedure. $G_{1}$ (main) and $G_{2}$ are generators while $D_{1}$ and $D_{2}$ are discriminators. Herein, $\mathbf{z}_{1,i} \in p_{z,1} \sim \mathcal{N}(0,I)$ and $\mathbf{z}_{2,i} \in p_{z,2} \sim \mathcal{N}(0,0.4I)$. Additionally, $\mathbf{z}_{1,i}^{c}$ and $\mathbf{z}_{i}^{*}$ indicate the candidate variable and the optimized safe vector, respectively.} 
  \label{fig:overview-defenseCycle}
  \vspace{-15pt}
\end{figure*}

The fourth variant of Eq.~\ref{eq:generalEqu}, proposed by Chen {\it et al.}~\cite{chen2020metamorph}, is called Metamorph. The EOT operation incorporated in this attack is based on a novel filter set using channel impulse response (CIR) operators. CIR is fundamentally similar to the RIR but instead of only simulating room configurations, it mainly focuses on the speaker-microphone (SM) pairs' geometrical position:
\begin{equation}
    \min_{\delta} \alpha_{m} l_{\text{dB}}(\vec{x}_{\mathrm{adv}})+\frac{1}{m}\mathcal{L}(\vec{x}_{\mathrm{adv}},\hat{\mathbf{y}}_{i}) 
    \label{eq:metamorphattack}
\end{equation}
\noindent where $\alpha_{m}$ makes a trade-off between the adversarial signal quality and the attack success rate, and $m$ refers to the total number of SM pairs. This attack was developed for the DeepSpeech model, and it has shown an outstanding performance in debasing the transcription performance of such a speech-to-text system. Qin {\it et al.}~\cite{qin2019imperceptible} developed the fifth variant of Eq.~\ref{eq:generalEqu}. They introduced a very reliable implementation for the EOT operation, which is called the Robust Attack. Moreover, this white-box attack is targeted and uses both $\ell_{net} (\cdot)$ and a masking threshold loss function $\ell_{m} (\cdot)$ as follows:
\begin{equation}
    \min_{\delta}\mathbb{E}_{t\sim \tau_{c}}\left [ \ell_{net}\left ( \mathbf{y}_{i},\hat{\mathbf{y}}_{i} \right )+c_{i} \ell_{m}(\vec{x}_{\mathrm{orig},i},\delta_{i}) \right ]
    \label{eq:robustatt}
\end{equation}
\noindent where $\tau_{c}$ is the filter set defined after CIR simulations. The Robust Attack was developed to attack the Lingvo transcription system, and the experiments have demonstrated the capability of this algorithm in crafting high-quality adversarial signals. Developing a black-box variant for Eq.~\ref{eq:generalEqu} is challenging since simulating RIR and CIR filter sets using common environmental settings might not be feasible. However, there are some approximation-based attack algorithms, such as the multi-objective optimization attack (MOOA)~\cite{khare2018adversarial} and the genetic algorithm attack~\cite{taori2019targeted}. These attacks are based on building a surrogate model for the victim transcription system via heuristic or greedy formulation. 

This paper proposes a defense approach for counteracting the adversarial attack algorithms mentioned above. In summary, this paper makes the following contributions: (i) an adversarial defense algorithm based on a cyclic GAN adapted for 2D wavelet representation; (ii) novel architectures for generator and discriminator networks for improving stability; (iii) characterization of the effectiveness of our defense approach against white and black-box adversarial attacks; (iv) evaluating the impact of defense algorithms on the quality of the signals.

\section{Background: Adversarial Defense}
\label{sec:background}
The algorithms for defending transcription systems against adversarial attacks fit the reactive defense category to the best of our knowledge. Sallo {\it et al.}~\cite{sallo2020adversarially} proposed the only proactive defense by adversarial training for short signals. Generally, the reactive defense algorithms can be categorized into explicit and implicit subcategories. The former subcategory includes algorithms that run filtration operations directly on the given input speech signal to bypass (modulate) the potential adversarial perturbation. For instance, MP3 encoding and multi-rate compression \cite{das2018adagio} have been employed for modulating adversarial signals. Das {\it et al.}~\cite{das2017keeping} inspired these defense approaches and they shown the positive impact of the low-level signal compression for bypassing the adversarial perturbation. However, a similar reactive approach with a high-level signal modulation perspective has been proposed by Latif {\it et al.}~\cite{latif2018adversarial}. This defense algorithm employs an autoencoder-based GAN (A-GAN) for reconstructing features of the given speech signal. However, it has been proven that both these two straightforward reactive approaches might not be able to bypass strong adversarial signals carefully crafted in enclosed environmental scenes~\cite{esmaeilpour2020class}. Additionally, the A-GAN suffers from instability during training, i.e., exploding gradient vectors after a high number of iterations \cite{brock2018large}, which results in low generalization of the generative model~\cite{esmaeilpour2020class}.

The second subcategory of reactive defense approaches includes algorithms that, instead of low or high-level filtrations, synthesize a signal very similar to the given input speech. These approaches are inspired by Samangouei {\it et al.}~\cite{samangouei2018defensegan}, and they implicitly avoid potential adversarial perturbation. During the last years, generative adversarial networks (GANs), such as multi-discriminator Mel-GAN \cite{kumar2019melgan} and class-conditional GAN \cite{esmaeilpour2020class}, have become reliable approaches for signal synthesis. The latter generative model has been mainly developed for adversarial defense purposes and uses shared embeddings with multiple sequential linear and residual blocks. This approach, called class-conditional defense GAN (CC-DGAN), iteratively finds a safe input vector ($\mathbf{z}_{i}^{*}$) for the generator network via:
\begin{equation}
\mathbf{z}_{i}^{*} \leftarrow \arg \min_{\mathbf{z}_{i} \in Z_{k}} \left \| \gamma\left [ G(\mathbf{z}_{i}),\mathbf{x}_{i} \right ] \right \|_{2}^{2}
\label{eq:CCDGAN}
\end{equation}
\noindent where $\gamma[\cdot]$ is an adjustment operator for measuring the distance between original and adversarial signal subspaces \cite{esmaeilpour2020detection,van1983matrix}. $G(\cdot)$ and $\mathbf{z}_{i} \in \mathbb{R}^{d_{z}}$ denote the generator network and the random latent variable with dimension $d_{z}$. Moreover, $\mathbf{x}_{i}$ refers to the discrete wavelet transform (DWT) spectrogram representation according to the settings mentioned in~\cite{esmaeilpour2020unsupervised}. Finally, running an inverse DWT operation on $G(\mathbf{z}_{i}^{*})$ reconstructs a high-quality signal that sounds like the input signal $\mathbf{x}_{i}$. This defense approach has been successfully tested against the adversarial attacks mentioned in Section~\ref{sec:intro}, but with a fairly lower model stability during training the generator network.
\section{Cyclic Defense GAN (CD-GAN)}
\label{sec:proposed}
We propose a more straightforward GAN-based defense approach for tackling the two major technical issues highlighted in the previous section. This novel implicit reactive adversarial defense approach is based on a cyclic GAN. It has three steps: converting a speech signal into a DWT spectrogram, finding a safe vector $\mathbf{z}_{i}^{*}$ for the cyclic generator network to synthesize a similar spectrogram, and reconstructing the speech signal with an inverse DWT operation.
\subsection{DWT Spectrogram}
Our motivation for generating DWT spectrograms rather than using 1D speech signals or using other 2D representation is threefold: spectrograms have much lower dimensionality and fit well with DNN architectures developed for computer vision applications; DWT most likely outperforms short-time Fourier transform in terms of providing distinctive features for GANs \cite{esmaeilpour2020unsupervised}; higher stability of the GAN during training \cite{esmaeilpour2020class}. Assuming that $a[n]$ is a discrete signal of length $n$, its DWT can be written as:
\begin{equation}
\mathrm{DWT}[\varrho ,n]=2^{\varrho /2}\sum_{\rho = 0}^{n-1}a[\rho]\psi[2^{\varrho},\rho-n]
\label{eq:DWT1}
\end{equation}
\noindent where $\varrho$ and $\rho$ denote the scale and dilation hyperparameters, respectively. Moreover, $\psi$ is the wavelet mother function, which is the complex Morlet function~\cite{young2012wavelet}. For obtaining the DWT spectrogram, we compute the power spectrum of this transformation as of $\mathrm{sp}_{\mathrm{DWT}}= \left | \mathrm{DWT}[\varrho ,n] \right |^{2}$.

\begin{table*}[t]
\footnotesize
\centering
\scriptsize
\caption{Performance comparison of defense approaches. Herein, the maximum number of iterations before complete collapse onsets are shown and modes are computed according to \cite{che2016mode}. Herein, E and I stand for Explicit and Implicit, respectively.}
\begin{tabular}{c||c||r|c c|c|c|c|c|c|c}
\hline
Model                        & Attack                         & \multicolumn{1}{c}{Defense}  & \begin{tabular}[c]{@{}c@{}}Iteration\\ ($\times 10,000$)\end{tabular} & \begin{tabular}[c]{@{}c@{}}Modes\\ $(\times 12.5)$\end{tabular} & \begin{tabular}[c]{@{}c@{}}Reactive\\ Type\end{tabular} & \begin{tabular}[c]{@{}c@{}}WER\\ (\%)\end{tabular}           & \begin{tabular}[c]{@{}c@{}}SLA\\ (\%)\end{tabular}             & STOI            & LLR & segSNR    \\ \hline \hline
\multirow{12}{*}{DeepSpeech} & \multirow{3}{*}{C\&W}          & A-GAN  & $01.59$                                                           & $0.89$                                                        & E   & $29.18\pm 2.1$  & $31.63 \pm 2.1$ & $\mathbf{0.84}$          & $\mathbf{0.41}$ & $18.42$ \\ \cline{3-11} 
                             &                                & CC-DGAN  & $02.67$                                                           & $2.55$                                                        & I   & $16.75 \pm 3.5$ & $60.17 \pm 1.2$ & $0.83$          & $0.37$ & $15.87$ \\ \cline{3-11} 
                             &                                & \textbf{CD-GAN}  & $\mathbf{02.91}$                                                           & $\mathbf{4.52}$                                                        & I   & $\mathbf{08.19 \pm 1.3}$ & $\mathbf{71.19 \pm 2.3}$ & $0.82$          & $0.44$ & $\mathbf{23.93}$ \\ \cline{2-11} 
                             & \multirow{3}{*}{Yakura's}      & A-GAN   & $01.22$                                                           & $0.66$                                                        &   E   & $20.57 \pm 0.6$ & $41.36 \pm 0.4$ & $0.83$          & $0.35$ & $21.06$ \\ \cline{3-11} 
                             &                                & CC-DGAN & $02.55$                                                           & $3.05$                                                        &  I   & $15.97 \pm 1.4$ & $62.19 \pm 1.2$ & $\mathbf{0.91}$          & $\mathbf{0.32}$ & $24.11$ \\ \cline{3-11} 
                             &                                & \textbf{CD-GAN}  & $\mathbf{02.61}$                                                           & $\mathbf{4.87}$                                                        & I   & $\mathbf{11.52 \pm 1.3}$ & $\mathbf{73.11 \pm 2.5}$ & $0.89$ & $0.34$ & $\mathbf{27.31}$ \\ \cline{2-11} 
                             & \multirow{3}{*}{Metamorph}     & A-GAN    & $01.04$                                                           & $0.71$                                                        &  E   & $19.97 \pm 1.7$ & $56.34 \pm 2.6$ & $0.92$          & $\mathbf{0.35}$ & $22.03$ \\ \cline{3-11} 
                             &                                & CC-DGAN & $\mathbf{02.98}$                                                           & $\mathbf{3.18}$                                                        & I   & $\mathbf{10.26 \pm 2.6}$ & $\mathbf{74.64 \pm 2.8}$ & $0.90$          & $0.36$ & $21.87$ \\ \cline{3-11} 
                             &                                & \textbf{CD-GAN}  & $02.91$                                                           & $2.55$                                                        & I   & $17.42 \pm 1.1$ & $70.82 \pm 2.3$ & $\mathbf{0.94}$          & $0.41$ & $\mathbf{25.18}$ \\ \cline{2-11} 
                             & \multirow{3}{*}{MOOA}          & A-GAN    & $01.27$                                                           & $0.54$                                                        &  E   & $19.67 \pm 3.6$ & $50.98 \pm 3.1$ & $\mathbf{0.92}$          & $0.34$ & $22.73$ \\ \cline{3-11} 
                             &                                & CC-DGAN & $02.89$                                                           & $3.76$                                                        &  I   & $12.32 \pm 1.2$ & $62.71 \pm 3.5$ & $0.89$          & $\mathbf{0.30}$ & $\mathbf{26.98}$ \\ \cline{3-11} 
                             &                                & \textbf{CD-GAN}  & $\mathbf{02.94}$                                                           & $\mathbf{4.11}$                                                        & I   & $\mathbf{07.36 \pm 2.1}$ & $\mathbf{71.11 \pm 2.4}$ & $0.91$          & $0.35$ & $24.36$ \\ \hline \hline
\multirow{3}{*}{Kaldi}       & \multirow{3}{*}{Imperio}       & A-GAN   & $01.02$                                                           & $0.65$                                                        & E    & $19.58 \pm 1.3$ & $51.87 \pm 2.1$ & $0.94$          & $0.37$ & $21.58$ \\ \cline{3-11} 
                             &                                & CC-DGAN & $02.63$                                                           & $2.97$                                                        &  I   & $12.87 \pm 2.1$ & $62.99 \pm 1.3$ & $\mathbf{0.96}$          & $\mathbf{0.32}$ & $23.67$ \\ \cline{3-11} 
                             &                                & \textbf{CD-GAN}  & $\mathbf{02.75}$                                                           & $\mathbf{3.63}$                                                        & I  & $\mathbf{07.49 \pm 1.5}$ & $\mathbf{71.01 \pm 1.9}$ & $0.92$          & $0.34$ & $\mathbf{26.94}$ \\ \hline \hline
\multirow{3}{*}{Lingvo}      & \multirow{3}{*}{Robust Attack} & A-GAN    & $01.02$                                                           & $0.56$                                                        &  E   & $18.88 \pm 1.2$ & $58.54 \pm 1.6$ & $\mathbf{0.95}$          & $\mathbf{0.30}$ & $17.52$ \\ \cline{3-11} 
                             &                                & CC-DGAN & $02.95$                                                           & $2.77$                                                        &  I   & $11.51 \pm 2.3$ & $62.58 \pm 1.7$ & $0.91$          & $0.33$ & $19.05$ \\ \cline{3-11} 
                             &                                & \textbf{CD-GAN}  & $\mathbf{02.96}$                                                           & $\mathbf{3.29}$                                                        &  I   & $\mathbf{09.45 \pm 1.2}$ & $\mathbf{70.96 \pm 0.8}$ & $0.94$          & $0.34$ & $\mathbf{22.88}$ \\ \hline
\end{tabular}
\label{table:comparnn}
\vspace{-15pt}
\end{table*}
\subsection{Spectrogram Synthesis: Safe Vector Optimization}
An overview of the proposed algorithm toward achieving a safe vector for the main generator network ($G_{1}$) to produce spectrograms similar to $\mathrm{sp}_{\mathrm{DWT}}$ is depicted in Fig.~\ref{fig:overview-defenseCycle}. There are two generators ($G_{1}, G_{2}$) in a cyclic framework connected with two fully dependent discriminator networks ($D_{1}, D_{2}$). Unlike conventional cyclic GANs (e.g., \cite{zhu2017data}), we do not provide source and target inputs to the generators for mapping from one sample to another. Instead, we employ $G_{2}$ mainly as a regularizer for $G_{1}$ to tackle the stability and mode collapse issues, i.e., losing sample variation during synthesis~\cite{brock2018large}). Concerning the superior performance of the least-square GAN (LS-GAN) configuration among generative models with symmetric divergence metrics \cite{hong2019generative}, we opted for this configuration for both $G_{1}$ and $G_{2}$. However, we use different settings for these networks to avoid the potential oversmoothing issue~\cite{hong2019generative,mao2017least}:
\begin{equation}
    \min_{G_{j}}\frac{1}{2}\mathbb{E}_{\mathbf{z}_{j,i}\sim p_{z,j}}\left [ (D_{j}(G_{j}(\mathbf{z}_{j,i}))-\vartheta_{1})^{2} \right ], \forall j \in \left \{ 1,2 \right \}
    \label{eq:lsgan1}
\end{equation}
\noindent where $p_{z,j}$ denotes the random sample distributions. Moreover, we initialize $\vartheta_{1}$ to one and zero for $G_{1}$ and $G_{2}$, respectively, in compliance with the standard LS-GAN configuration \cite{hong2019generative}. We empirically designed slightly different architectures for these generators to make a reasonable trade-off between model generalizability and stability. The main generator contains six hidden layers: a fully connected (4$\times$4$\times$16 channels), two stacked residuals (with 16$\rightarrow$8 and 8$\rightarrow$4 channels plus 512 filters), and three consecutive convolution blocks (padded with receptive fields 5$\times$5$\times$1 plus 256 filters) followed by batch normalization and ReLU activation function. The output layer is a transposed convolution \cite{mao2018effectiveness} with $\tanh$, resulting in a 128$\times$128$\times$3 spectrogram. The second generator is simpler and contains three sequential 3$\times$3$\times$1 convolutional layers with 128 filters, skip-$z$ through these layers \cite{brock2018large}, and average pooling. The output layer of $G_{2}$ is a non-local layer with a 16$\rightarrow$4 channel and max-pooling operation. For training the discriminator networks, we also use the standard LS-GAN configuration, which iteratively minimizes for \cite{hong2019generative,mao2017least}:
\begin{multline}
   \quad \quad \quad \quad \quad \min_{D_{j}}\frac{1}{2}\mathbb{E}_{\mathbf{x}\sim p_{r}}\left [ D_{j}(\mathbf{x})-1)^{2} \right ]+\\ \mathbb{E}_{\mathbf{z}_{j,i} \sim p_{z,j}}\left [ (D_{j}(G_{j}(\mathbf{z}_{j,i}))-\vartheta_{2})^{2} \right ] , \quad \forall j \in \left \{ 1,2 \right \}
\end{multline}
\noindent where $p_{r}$ indicates the real sample distribution. Additionally, $\vartheta_{2}$ is $\left \langle 0,-1 \right \rangle$ for $D_{1}$ and $D_{2}$, respectively. For avoiding unnecessary complications and computational overhead, we use an identical architecture for both discriminator networks. This unique architecture requires a spectrogram with a dimension 128$\times$128$\times$3 in the input layer on the top of the five stacked hidden layers, namely two convolutions and three residuals. For the convolution blocks, we train 128 filters with receptive fields 3$\times$3$\times$1, followed by batch normalization and leaky ReLU activation function. On top of the residual blocks, containing 256 filters with 4$\rightarrow$4 and 4$ \rightarrow$1 channels, respectively, there is one non-local layer with 16 channels, max pooling, ReLU, and a linear logit layer ($\rightarrow$1). For training our cyclic GAN, we extend the cycle-consistency loss function introduced in \cite{esmaeilpour2020unsupervised} as:
\begin{equation}
    \mathcal{L}_{cyclic}(\cdot)=\mathcal{L}(G_{1},D_{2})+\mathcal{L}(G_{2},D_{1})+\alpha_{c}\mathcal{L}(G_{1},G_{2})
\end{equation}
\noindent where $0< \alpha_{c}\leq 1$ is the cyclic scaling coefficient that should be empirically tuned during training. However, for simplicity and reproducibility purposes, we set this hyperparameter to $0.9$. As shown in Fig.~\ref{fig:overview-defenseCycle}, we first minimize the dissimilarity between the input and the synthesized spectrograms (red rectangle) to achieve the candidate vector $\mathbf{z}_{1,i}^{c}$. This vector forces the main generator to yield a spectrogram similar to $\mathbf{x}_{i}$. We later refine this vector by minimizing the dissimilarity between the outputs of $G_{1}$ and $G_{2}$ (blue rectangle). Upon convergence of this minimization procedure, we achieve the safe vector $\mathbf{z}_{i}^{*}$ for synthesizing the final spectrogram.  

\subsection{Signal Reconstruction}
The last step of our adversarial defense approach is to reconstruct the speech signal from the synthesized spectrogram using the optimized safe vector. We use the main generator to craft $G_{1}(\mathbf{z}_{i}^{*}) \mapsto \mathrm{sp}_{\mathrm{DWT}}^{*}$. This spectrogram is very similar to the given input spectrogram $\mathbf{x}_{i}$ and does not carry the potential adversarial perturbation. For reconstructing the speech signal, we run the inverse DWT operation on the obtained $\mathrm{sp}_{\mathrm{DWT}}^{*}$~\cite{cvejic2002wavelet,masuyama2019deep,meyer1992wavelets}.

\section{Experiments}
\label{sec:experiments}
This section explains the conducted experiments on three cutting-edge transcription systems, namely DeepSpeech, Kaldi, and Lingo. These speech-to-text models are trained on MCV \cite{MozillaCommonVoiceDataset} and LibriSpeech \cite{panayotov2015librispeech} datasets, which contain short and long voice recordings. We randomly selected 15,000 English-speaking samples from these datasets, including different utterances from various ages and genders. We use 70\% of these samples for training the GANs and keep the remaining portion for developing adversarial attacks, as discussed in Section~\ref{sec:intro}. The main motivation for crafting adversarial signals only for a part of these datasets follows a common practice in adversarial attack development and analysis~\cite{carlini2018audio,qin2019imperceptible,chen2020metamorph,schonherr2020imperio,yakura2018robust,esmaeilpour2020class}. Furthermore, the proposed defense approach does not depend on the amount of benchmarking samples.

We converted the training speech signal into $\mathrm{sp}_{\mathrm{DWT}}$
for training our cyclic GAN. We set the DWT sampling rate to 16 kHz with a frame length of 50 ms and an overlapping ratio of 0.5. Finally, we rescaled all the spectrogram to 128$\times$128$\times$3, matching the input layers of the generator networks. We also make identical assumptions for the RIR, CIR, microphone-speaker position, and room settings for all the attack algorithms, as discussed in Section~\ref{sec:intro}. Moreover, we assign five incorrect phrases ($\hat{\mathbf{y}}_{i}$) to the targeted and non-targeted attacks (e.g. MOOA) randomly selected from the corresponding datasets. Finally, we compare the performance of the defense algorithms against these attacks using seven objective metrics in three categories: two metrics for measuring the defense success rate; three metrics for evaluating the quality of the signals after running defenses; two metrics for assessing the generalizability, and stability of the generative models.

For the first category, we implemented sentence-level accuracy (SLA) and word error rate (WER) as discussed in \cite{qin2019imperceptible}. According to the definitions of these metrics, a reliable defense approach should result in higher SLA and lower WER~\cite{qin2019imperceptible}. For the second category, we use segmental signal-to-noise-ratio (segSNR) \cite{baby2019sergan}, short-term objective intelligibility (STOI) \cite{taal2011algorithm}, and log-likelihood ratio (LLR) \cite{baby2019sergan} which measure the relative quality of the given signals regarding the environmental noises. The latter metric has an inverse relationship with the rest, and for a higher quality signal, the LLR is lower. Finally, for the third category, we employ the maximum number of iterations before complete collapse onset \cite{brock2018large} and a total number of learned modes \cite{miyato2018spectral} for a batch size of 2$\times$512. 

Table~\ref{table:comparnn} summarizes our achieved results averaged over ten repeating experiments. For most cases, the proposed adversarial defense approach (CD-GAN) outperforms other defense algorithms in terms of model stability (higher number of iterations before collapse onset and modes per batch) and defense success rate (lower WER and higher SLR). Thus, there is a direct relation between model stability and defense success rate. In other words, developing more stable models most likely yields a more reliable defense approach. But, on the other hand, our CD-GAN often marginally fails against other defenses according to STOI and LLR metrics in terms of the quality of the reconstructed signals. This is presumably due to a slight spectrogram oversmoothing side-effect, which usually happens at higher iterations \cite{esmaeilpour2020unsupervised}. Employing similar loss functions for the GANs and the non-local characteristic of the metrics mentioned above~\cite{baby2019sergan,taal2011algorithm} are the principal reasons for subtle differences among the reported values. For tackling this issue, we exploit the segSNR metric, which locally measures the quality of the speech signals. According to this metric, our proposed defense algorithm outperforms other approaches for the majority of the cases.    

\section{Conclusion}
This paper introduced a novel adversarial defense algorithm against cutting-edge speech adversarial attacks. Our defense approach is based on a cyclic GAN framework employing two generator and discriminator networks. These networks implement layers of convolution and residual blocks for capturing local and global distributions of DWT spectrograms. This procedure contributes to reconstructing a signal almost without adversarial perturbation. Although we have shown that our proposed CD-GAN outperforms other algorithms in terms of model stability and defense success rate, it might not produce noise-free signals. We demonstrated that, our proposed algorithm has competitive performance to other defense approaches using segSNR, STOI, and LLR quality metric. In our future work, we will employ some regularizers on the cycle-consistency loss function based on human psychoacoustic hearing thresholding to improve our defense algorithm in terms of signal preservation after reconstruction.

\balance

\bibliographystyle{IEEEtran}

\end{document}